\documentclass[pdflatex,10pt,JHEP.bst]{article}
\usepackage{hyperref}%
\usepackage{graphicx}%
\usepackage{amsmath,amssymb,amsfonts}%
\usepackage{mathrsfs}%
\usepackage{xcolor}%
\usepackage{manyfoot}%
\newcommand{\al}{\alpha}
\newcommand{\bt}{\beta}
\newcommand{\gm}{\gamma}

\newcommand{\kp}{\kappa}

\newcommand{\rh}{\rho}
\newcommand{\sg}{\sigma}

\newcommand{\ch}{\chi}
\newcommand{\ps}{\psi}
\newcommand{\psb}{\bar\psi}

\newcommand{\dmu}{\partial_\mu}
\newcommand{\half}{\frac{1}{2}}
\newcommand{\intx}{\int d^4 x\,}
\newcommand{\Tr}{\mbox{Tr}}
\newcommand{\Qtop}{Q_{\rm top}}
\newcommand{\chtop}{\ch_{\rm top}}
\newcommand{\SF}{S_{\rm F}}
\newcommand{\Nf}{N_{\rm f}}
\newcommand{\Mc}{M_{\rm cr}}

\begin{document}
\begin{center}{{\Large\bf A confederacy of anomalies}\\[1cm]
\setcounter{footnote}{0}
\renewcommand{\thefootnote}{\fnsymbol{footnote}}
Jan Smit\footnote[1]{e-mail: j.smit@uva.nl}\\[5mm]
Institute for Theoretical Physics, Institute of Physics,\\
University of Amsterdam, Postbus 94485, 1090 GL Amsterdam, Netherlands\\[2cm]

{\bf Abstract}\\[5mm]
A personal recollection of early years in lattice gauge theory with a bias towards chiral symmetry and lattice fermions.\footnote[2]{
Extended write-up of a talk at the 41st International Symposium on Lattice Field Theory {\em Lattice 2024}, 
28 July - 3 August 2024, Liverpool, UK.}\\[2cm]

\setcounter{footnote}{0}
\renewcommand{\thefootnote}{\arabic{footnote}}
}

\end{center}

\section{Prologue}

September 1969, at the Institute for Theoretical Physics in Amsterdam (ITFA), I became a PhD student of S.A.\ (Sieg) Wouthuysen. 
Recent developments in the field of high-energy physics (HEP) had his attention somewhat from a distance, for his own research he was more interested in fundamentals. Sieg's students were free to choose research topics and he was very helpful supporting them. 

In 1970 I was working on my own while also trying to understand incoming HEP literature with fellow student G.J.\ (Gerbrand) Komen\footnote{Gerbrand spent a year at the  International Centre of Theoretical Physics in Trieste, defended his PhD thesis in Amsterdam in 1972, and after occupying postdoc positions (CERN, Leiden University) he changed to the Royal Dutch Meteorological Institute (KNMI), became eventually a director there and an associated professor at Utrecht University.}.
Information about what was going on in those days came from several sources. In the easily accessible library one eagerly perused a weekly pile of new preprints and also new issues of journals. Inevitably this implied a delay in the stream of current developments. Topics enough: Regge poles, current algebras, effective Lagrangians, partons, scale and conformal invariance, renormalization group \ldots 
With Gerbrand I studied deep inelastic scattering, partons and even conformal invariance.
Another stimulating source was the National Seminar instigated by M.J.G.\ Veltman (Utrecht University) and organised in turn about every six weeks by the HEP theory groups of the Dutch universities. Veltman's focus on massive vector bosons for the weak interactions \cite{Veltman:1968ki} got us interested in the work of Glashow, Weinberg, Salam and Yang-Mills fields. 
C.P.\ (Chris) Korthals Altes\footnote{Chris obtained his PhD in Amsterdam in 1969 and had moved to Oxford for his first postdoc position. He settled at the Centre de Physique Th\'{e}orique, CNRS in Marseille.}   suggested to me the problem of renormalizability of Weinberg's model for leptons \cite{Weinberg:1967tq}. I tried, but got stuck in the unitary gauge with its strongly UV divergent Feynman diagrams, like so many others. 
The presentation by Veltman's student Gerard 't Hooft at the 1971 HEP conference in Amsterdam showing renormalizability was a big surprise also for us students. After this event the quantum field theory approach to the weak interactions and to particle physics in general got a great boost.

\emph{My hero} was Kenneth G.\ Wilson, because of his invention of the operator product expansion \cite{Wilson:1969zs,Wilson:1970pq} and his treatment of the renormalization group (RG) \cite{Wilson:1970ag,Wilson:1971bg,Wilson:1971dh}. His work on the Thirring Model \cite{Wilson:1970pq} inspired me to write an article on conformal invariance 
\cite{Smit:1971zz}\footnote{It was submitted to {\em Communications in Mathematical Physics} 
and judged unsuitable for the journal. A loner lacking experience in such situations, I did not pursue getting it published elsewhere.}.
In  \cite{Wilson:1971dh} Wilson evaluated a path integral by approximating its field configurations in terms of wave packets localised in space and wave-number (`momentum') space, and used this in formulating a discrete RG transformation. I studied  this work together with F.W.\ (Frits)  Wiegel\footnote{Frits had left industry for academia and obtained his PhD in 1973 at the ITFA on the thesis \emph{Some applications of functional integration to classical and Bose condensation}. He became professor at Twente University (Enschede) in 1975.}, then already an aficionado of using path integrals in condensed matter physics \cite{Wiegel:1977kp}.
No longer the path integral was just a magic trick for deriving Feynman diagrams, it had become a concrete existing object that could be approximated non-perturbatively.\footnote{In those days even the derivation of Feynman rules with massive vector particle was not straightforward and Boulware's publication \cite{Boulware:1970zc} showed that the path-integral method gave results easily and quickly.  
Initially field theoretic path integrals, including fermionic ones, came to me 
through the book by J.\ Rzewuski \cite{Rzewuski}, picked up by chance in a famous bookstore in London.}

In 1972 two articles appeared stating that chiral anomalies \cite{Adler:1969gk,Bell:1969ts} 
should cancel in gauge theories \cite{Bouchiat:1972iq,Gross:1972pv}.
This struck me as artificial\footnote{I had missed Adler's lucid Brandeis lectures \cite{Adler:1970Brandeis} and did not understand well enough the inevitability of anomalies in electro-weak models such as Weinberg's \cite{Weinberg:1967tq}.} since the anomalies had to do with violations of Ward-Takahashi identities due to UV divergent Feynman integrals. These identities could conveniently be derived by transformations of variables in path integrals. However, in case of anomalies the path integral could still not be trusted. I felt the need of an unambiguous non-perturbative regularization of path integrals such that identities obtained by transformations of variables were simply true. Wilson's localized wave packets in \cite{Wilson:1971dh} somehow suggested trying a lattice.   

Putting the scalar field on a lattice (Spring 1972) felt like a kind of blasphemy in giving up Lorentz invariance and it gave a thrill getting it back in the continuum limit. Implementing gauge invariance had to wait a while. That Summer  R.J. (Bob) Finkelstein visited the ITFA. Bob and Sieg were students of J.R.\ Oppenheimer and long time friends. The idea came up that I become Bob's student at the University of California, Los Angeles (UCLA). Some guidance by a practitioner of theoretical HEP (Bob) should provide me with the focus needed for publishing papers.\footnote{
Teaching and Research Assistantships at UCLA and the Dutch Science Foundation FOM gave financial support.}

August 1972 I moved to LA and started working with Bob ``hands on'', on the triple vertex function of massive Yang-Mills fields, in Schwinger's Source Theory \cite{Schwinger:1969vz,Schwinger:1970xc,Schwinger:1973rv}. In the language of Feynman diagrams the massive Yang-Mills model showed signs of renormalizability at one loop \cite{Veltman:1968ki}, but not at higher loop orders \cite{Boulware:1970zc}, and Bob was interested to see how the methods of source theory would fare in this problem.\footnote{The relatively new source theory bypassed working with ill-defined quantities occurring in quantum field theory practice. At UCLA, Julian Schwinger had three collaborators
producing papers on applications of source theory. It appeared to me elegant and useful at the level of effective Lagrangians.}

However, I could not refrain from trying to find a lattice regularization of non-Abelian gauge field theory. At Christmas 1972 the problem was solved at the kitchen table when I arrived at the lattice action using unitary gauge-group fields (generally known as the Wilson action). 
In 1973 I included fermions in the ``naive'' way and investigated the perturbative continuum limit at one loop in a $U(1)_V\times U(1)_A$ gauge-Higgs model coupled to one Dirac field, with emphasis on chiral anomaly aspects. The fermion part of the Larangian was given by   
\[
\mathcal{L}_{\rm F} = \psb\gm^\mu\partial_\mu\ps + \psb\left[g\,\gm^\mu\,V_\mu +g_5\, i\gm^\mu\gm_5\,A_\mu +G(\,\sg+ i \gm_5\,\pi) \right]\ps\,.
\]
It is invariant under vector ($\al$) and chiral ($\bt$) gauge transformations
\[
\ps\to e^{i\al + i \bt \gm_5}\,\ps\,,\quad 
\psb\to \psb e^{-i\al + i \bt \gm_5}\,\ps\,,\quad 
\sg + i \pi\to e^{-2 i \bt}\, (\sg + i \pi)\,.
\]
The remainder of the Lagrangian depended only on the gauge fields $V_\mu$, $A_\mu$ and the complex Higgs field $\sg+i\pi$. When the Higgs field gets a real vacuum expectation value, $\langle\sg\rangle$, chiral symmetry becomes broken and the fermions get a mass $m=G\langle \sg \rangle$.

The model was put on a spatial cubic lattice with lattice spacing $a$ of length $L = N\, a$ on the side, integer $N>0$. Time was kept continuous (just like quantum mechanics). Mimicking continuum procedures I calculated one-loop vector $V$ and $A$ selfenergy (`vacuum polarization') diagrams and fermion $VVA$ and $VV\pi$ triangle diagrams. Ward-Takahashi relations (\emph{in particular chiral ones}) were valid at finite lattice spacing $a$, also in the infinite volume limit $N\to\infty$. 
After renormalization a continuum limit $a\to 0$ could be taken. The limiting diagrams had the usual Lorentz-invariant form, \emph{but the $V$ and $A$ selfenergies were a factor 8 too large and the triangle diagrams came out zero!}

\begin{figure}
\includegraphics[width=7cm]{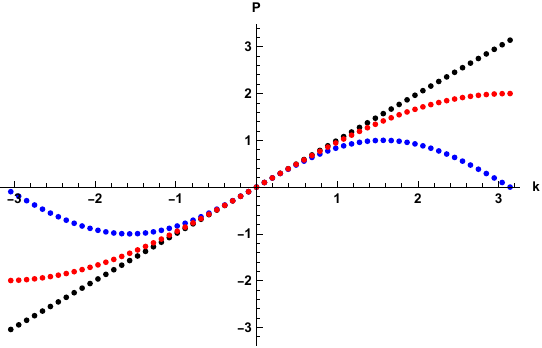}
\includegraphics[width=7cm]{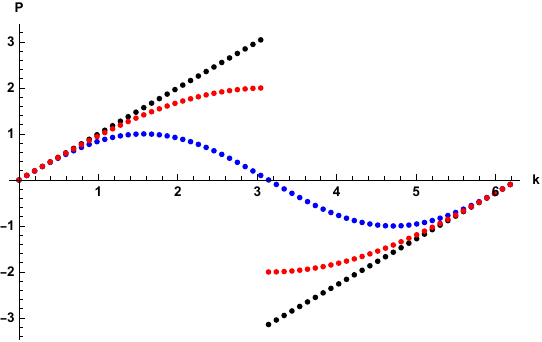}
\caption{Momentum functions $P_j(k)$ of an $N=64$ lattice versus $a k_j$ in the interval $(-\pi,\pi]$. Bottom to top at $a k_j=2$:  $P_j(k) = \sin(ak_j)/a$ (``naive''),  $2 \sin(a k_j/2)/a$ (``modified''), $k_j$ (``SLAC''). 
In the right plot, the left-plot modes in $(-\pi, 0)$ have been shifted by $2\pi$: $0\leq a k_j <2\pi$. 
 Lattice units $a=1$.
}
\label{figP}
\end{figure}

The factor 8 could be understood by looking at the fermion propagator in momentum space, $\SF(k)$:
\[
\SF(k)^{-1} = m+i \gm^0 k_0 + \sum_{j=1}^3 i \gm^j P_j(k)\,,\quad
P_j(k) = \frac{1}{a}\, \sin(a k_j)\,.
\]
The $j$-component of the three-vector $\vec{P}(k)$ is illustrated in the left plot of Figure \ref{figP} by the blue dots (lowest at $a k_j=2$). 
The lattice propagators and vertex functions are periodic functions of the spacial momenta $\vec{k}$.
(There is an equal number of $\vec{k}$ values and $\vec{x}$ values.)
In the continuum limit only the infinitesimal neighbourhood of the zeros of the vector function $a \vec{P}(k)$, i.e.\ $a k_j=\{0,\pi\}$, $j=1,2,3$, or
\[ 
a\vec{k}=(0,0,0), (\pi,0,0), (0,\pi,0), (0,0,\pi), (\pi,\pi,0), (\pi,0,\pi), (0,\pi,\pi), (\pi,\pi,\pi)\,,
\]
give non-polynomial contributions in the Feynman diagrams, $8$ in total. The rest of the momentum integration gives polynomial terms absorbed by the renormalization procedure.

I tried to cure the factor 8 problem by choosing the modified function
\[
P_j(k) = \frac{2}{a}\, \sin\left(\frac{a k_j}{2}\right)\,.
\]
As the right plot in Figure \ref{figP} shows at $a k_j=\pi$, this function is discontinuous on the periodic momentum space. Consequently its Fourier transform to position space, the corresponding lattice gradient operator, is non-local. 
It falls off $\propto |x_j-y_j|^{-1}$ and for  gauge invariance this has to be accompanied by a product of the unitary gauge-field link-variables between lattice sites $x$ and $y$. Investigating the effect of the discontinuity in the continuum limit, I found non-local UV divergent terms violating Lorentz invariance which could not be 
eliminated by the renormalization procedure. This was obviously unacceptable.

The fermion doubling problem, i.e.\ the non-zero contribution of the regions near $a k_j = \pi$, remained unsolved. I was not sure the `doublers' were to be considered as physical particles, could they be produced in the laboratory? All 8 particles would have the same mass $m$ and 7 copies of the electron did not exist in Nature. The relevance of the triangle diagrams being zero was also a mystery. I could have talked to Bob about these issues, but did not in sufficient detail and being stuck decided to leave the lattice alone for a while. 

November 1973 the Source Theory article with Bob was ready \cite{Finkelstein:1974xm}. Inspired by a paper of Cornwall and Norton at UCLA on spontaneous symmetry breaking in theories \emph{without} elementary scalar fields 
using Schwinger-Dyson equations 
\cite{Cornwall:1973ts}, I began investigating such a possibility in the Yang-Mills model, which led to \cite{Smit:1974je}.

Information about what was going on in particle physics was somehow less accessible to me in LA than in Amsterdam.\footnote{But I noticed the influential review of Kogut and Wilson \cite{Wilson:1973jj} in which continuous scalar fields were used for RG transformations in the discrete Ising model.} Spring/Summer 1974 Wilson gave a seminar at UCLA that hit me very hard. My hero talked about the lattice, the same formulation as mine, with wonderful concepts and results \cite{Wilson:1974sk}: 
\[
\mbox{strong coupling!\quad  Wilson loop!\quad confinement!}
\]
After the talk I went up to him and mentioned the problem with fermions (but did \emph{not} tell him I had also `invented' lattice gauge theory). He seemed not surprised  and said: {\em Add a term 
$\propto\dmu\psb\dmu\ps$ to the action}. Something I had not dared to think about because it breaks chiral symmetry. 
Starting my {\em But} \ldots he added: {\em Chiral symmetry should come back in the continuum limit}.\footnote{I got the impression that people had already been digesting his work. Article \cite{Wilson:1974sk} (submitted 12 June 1974) referred to upcoming publications by Kogut and Susskind, by Balian, Drouffe and Itzykson (who refer to a seminar at Orsay in 1973), and to a conference at Marseille in 1974.}

December 1974 the PhD defense took place\footnote{Exams of regular graduate courses were waived, having taken similar ones in Amsterdam, except the required `final comprehensive'. Preparation for it was still very nice and it was a also pleasure attending lectures by Finkelstein, Sakurai and Schwinger.}, on the thesis: \emph{Massive Vector Particles with Yang-Mills Couplings}.
It contains further work with the source theory method and article \cite{Smit:1974je} using Schwinger-Dyson equations. Lattice regularization was only briefly mentioned.
January 1975 I returned to Amsterdam, to a position of the Dutch Science Foundation.
Depressed by not seeing sufficient reason for writing up my lattice results after Wilson's great talk and paper \cite{Wilson:1974sk} {\em and solving the doubling problem}, I wished to do research relevant to experimental colleagues at the Zeeman Laboratory who were involved in hadron scattering experiments at CERN. This led to working on Regge pole theory. Backward pion-nucleon scattering was tackled by summing infinite `towers' of nucleon \& delta resonance exchanges. 
A new PhD student joined, L.H.\ (Luuk) Karsten. Comparison with real experimental results turned out give a special exiting thrill.

\section{Back to the lattice}

September 1977 I attended a SLAC Summer Institute where Sidney Drell lectured about fermions and gauge fields on the lattice. He proposed the fermion method presented in the paper together with Weinstein and Yankielowicz \cite{Drell:1976mj}, the DWY or SLAC derivative:
\[P_j(k) = k_j \,,
\]
as a solution to the doubling problem while keeping `local $\gm_5$ invariance'. The similarity with the ``modified'' $P_j(k)$ in Figure 1 suggested to me problems in the continuum limit also with his proposal. 
After the lecture I mentioned to Drell having done perturbative weak coupling calculations in a lattice gauge theory using a derivative with a similar discontinuity, and that one should expect non-local UV divergencies breaking Lorentz invariance starting, at one loop. After some discussion he said: \emph{You ought to write this up !}

This seemed good advice, making it worthwhile to use the knowledge acquired in 1973  and do a one-loop calculation on a Euclidean lattice\footnote{Lattice in space and imaginary time (also used by Wilson). Imaginary time improves symmetry between space and time and convergence of path integrals. Monte Carlo computations use Euclidean lattices. }
 with $P_\mu(k) = k_\mu$. Luuk was happy to join the project, he had followed lectures by Susskind at the 1976 St.\ Andrews Summer Institute. 
We had to catch up with an exponentially rising number of publications in lattice gauge theory since Wilson's 1974 paper \cite{Wilson:1974sk}, in particular:\\[1mm]
- Wilson's lectures in which his fermion method appeared \cite{Wilson:1975id};\\
- Hamiltonian formulation of QCD by Kogut and Susskind \cite{Kogut:1974ag};\\
- Susskind's (staggered) fermion method \cite{Susskind:1976jm};\\
- strong coupling expansions 
by Balian, Drouffe and Itzykson \cite{Balian:1974xw};\\
- hadron spectrum by Banks, Raby, Susskind, Kogut, Jones, Scharbach, Sinclair\cite{Banks:1976ia};\\
- Sharatchandra's article on weak coupling perturbation theory in QED \cite{Sharatchandra:1976af};\\
- Baaquie's article on the non-Abelian vector selfenergy \cite{Baaquie:1977hz}.\\[1mm]
(I overlooked work on the transfer matrix by Creutz \cite{Creutz:1976ch} and by L\"{u}scher \cite{Luscher:1976ms}.)\\

Our first paper (March 1978) on the DWY fermion method concerned the $VVA$ triangle diagram in which we found indeed non-local Lorentz symmetry violating contributions in the continuum limit \cite{Karsten:1978nb}. The calculation was fairly involved and one might wonder what happens in models with anomaly cancelation. Lacking the response we had hoped for we decided to calculate also the vector selfenergy, which might be more convincing. It appeared a year later (May 1979) showing again unacceptable contributions in the continuum limit \cite{Karsten:1979wh}. 

Almost simultaneously with \cite{Karsten:1978nb} Shigemitsu's paper \cite{Shigemitsu:1978br} appeared using `Wilson fermions' in a strong coupling calculation. The large pion mass result of \cite{Banks:1976ia}  with staggered fermions 
did not occur with Wilson fermions in which $m_\pi/m_N$ could be tuned to zero while hardly affecting other mass ratios.
By 1979 lattice gauge theory had been used for diverse subjects as the ground state Yang-Mills wave functional \cite{Greensite:1979ha}, duality transformations \cite{Savit:1979ny}, Higgs phases \cite{Banks:1977cc,Fradkin:1978dv}, and the confinement mechanism beyond strong coupling \cite{Mack:1978kr,Munster:1979pg}.

September 1979 at a Carg\`{e}se Summer Institute which I attended, Wilson described his RG results using  Monte Carlo computations \cite{Wilson:1979wp}. He mentioned calculations of the QCD RG $\beta$-function connecting strong to weak coupling by Kogut, Pearson and Shigemitsu \cite{Kogut:1979vg}, and also Monte Carlo computations by Creutz, Jacobs and Rebbi \cite{Creutz:1979zg}, and by Creutz \cite{Creutz:1979cu,Creutz:1980zw}.
The latter (followed by \cite{Creutz:1980wj}) showed a string tension following weak coupling RG behavior. 
This was an \emph{eye opener}: it demonstrated that Monte Carlo methods were able to compute {\em nonperturbative} observables of QCD at {\em weak coupling} where continuum behaviour should take place.

October 1979 Luuk obtained his PhD degree after defending his thesis:
\emph{On Lattice Gauge Theories and On Backward Pion-Nucleon Scattering}.
The second part referred to the Regge pole work. In the first part he clarified the relation of fermion doubling to the chiral anomaly. 

Luuk used the more convenient four-dimensional Euclidean hypercubic lattice. In the fermion propagator this leads to $P_j(k) \to P_\mu(k) = \sin(ak_\mu)/a$, where the time-like component $k_4$ takes similar values as the spatial $k_j$, $j=1,2,3$. 
He showed that the doubler fermions are related by unitary transformations to the `wanted one'. Thus the doublers have to be viewed as physical particles. In lattice QED they can be pair-produced in electron-positron scattering. In the lattice  $U(1)_V\times U(1)_A$ model they all have the same vector charge $g$ but the unitary transformations show that axial charges may have a different sign. Representing this as $g_5\to g_5\,Q_5$, $Q_5=\pm 1$, the $Q_5$ are given by\footnote{With continuous time the sum of the axial charges reads: 
$\sum Q_5 = 1-3+3-1=0$. In the Euclidean lattice case $\sum Q_5 = 1-4+6-4+1=0$.}:\\[2mm]
$Q_5 = (-1)^{n} $,\;\; $n=0,1,\ldots,4$, the number of $\pi$s in the zero of $P_\mu(k)$,
\;\; $\sum Q_5=0$.\\[2mm]
The one-loop VVA triangle diagrams are linear in the axial charges and since all particles have the same mass and vector charge, the continuum limit of the lattice triangle diagrams is proportional to $\sum Q_5=0$. 
This explained why I found the vanishing result.

Furthermore, in the $U(1)_V\times U(1)_A$ Higgs model, a Wilson-type mass term made gauge invariant with the Higgs field $\sg + i \pi$ and parametrized by $G_{\rm W}$
raised the doubler masses in the continuum limit, giving the mass spectrum
\[
m_{\rm F}= m + 2 n\, \tilde r\,,\quad 
m=G\langle \sg \rangle\,, \quad \tilde r=G_{\rm W} \langle \sg\rangle\,.
\]
(Wilson's $\tilde r$ equals $1/a$.)\footnote{For QCD, Wilson added the lattice version of $(a r/2) \partial_\mu\psb\,\partial_\mu\ps$ to the lattice Lagrangian, choosing $r=1$. With this choice the doublers are eliminated even at finite lattice spacing.}
In the chiral Ward-Takahashi identity, the sum of the triangle diagrams is no longer zero. However, the anomalies are mass-independent and their sum $\propto \sum Q_5$ is still zero.
\emph{Chiral anomalies canceled in the triangle diagrams !}

Subsequently, the doublers ($n\geq 1$) could perhaps be given large unobservable masses. They might even be removable by sending $\tilde r\to\infty$, \emph{except} for triangle diagrams: their contribution becomes the chiral anomaly. 
But this limit was doubtful, $\tilde r \to \infty$ would imply $G_{\rm W} \to \infty$ and one would be led to problematic strong coupling dynamics. 

Luuk presented his work also at the August 1979 Kaiserslautern Study Institute  \cite{Karsten:1979qs}.
November 1979 he moved to Stanford. We still had to publish results in a more widely read journal and extend the calculations to QCD with Wilson fermions at fixed $r=a \tilde r$ instead of fixed $\tilde r$. 

That Autumn a new PhD student, Jaap Hoek, started  research on the computation of the effective action of {\em hadrons} in QCD using a strong coupling expansion with Wilson fermions.
1979/1980  Banks and Casher published a stimulating paper on chiral symmetry breaking  in which the chiral condensate was calculated in terms of the density of eigenvalues of the Dirac operator \cite{Banks:1979yr}.
July 1980, at a conference in Santa Barbara I met Noboru Kawamoto who became a postdoc in Amsterdam. He had just submitted an article on the phase structure of QCD with Wilson fermions \cite{Kawamoto:1980fd}. I talked about my paper \cite{Smit:1980nf} on chiral symmetry in QCD at \emph{strong coupling} with Wilson fermions. Some details:\\[1mm]
- Hamiltonian formulation,
$\Nf$ flavors, $N$ colors, order $1/g^2$, two orders in $1/N$;\\
\[ S_{\rm mass} = \psb (M -  W) \ps,\;\;\;\psb W\ps = \frac{a r}{2}\, \sum_{x,j} 
\left(\psb_xU_{j x}\ps_{x+a_j} + \psb_{x+a_j} U_{j x}^\dagger \ps_x \right)
\]
- critical mass $\Mc(g,r)\propto r^2$: 
choose $M\geq \Mc$ ($M<\Mc$ unphysical phase);\\
- at $r=0$: Nambu-Goldstone bosons (also the vector mesons $\rh$, $K^*$), $\langle\psb\ps\rangle \propto N$;\\
- raising $r$: for $r> 0.7$  vector mesons have lost their NGB character;\\
- but not the `charged' pseudoscalars ($\pi^\pm$, $K^\pm$, $K^0$, $\bar K^0$):\\
- `current quark mass' of flavor $a$:  $m_a=M_a-\Mc$, $m_{ab}^2 \propto (m_a + m_b)$, $a\neq b$ ;\\
- `neutral' $m_{aa}^2$ differ from $m_{ab}^2$ at next order in $1/N$ 
(\emph{$U(1)$ problem});\\
- vector meson `dynamical quark mass' $m_{\rm dyn}\simeq \Mc/2$,  
$m_{{\rm V} a b}\simeq 2 m_{\rm dyn} + m_a + m_b$;\\
- baryon masses depend strongly on $r$, $m_B \propto N$ at leading $N$. 

Thus, Wilson's fermion method showed dynamical chiral symmetry breaking in the `charged' meson mass spectrum even at strong coupling! A fit for $\Nf=3$ assuming $m_u=m_d$ gave acceptable values: $m_u\simeq 5$ MeV, $m_s \simeq 130$ MeV, $m_{\rm dyn} \simeq 380$ MeV.

Matrix elements of  currents (meson decay constants $f_\pi$, $\gm_{\rh}$, etc) did not come out very well (neither very badly). The shown currents were derived from a lattice version of the Standard Model that generated the $r$-parameter in terms of a `Wilson-Yukawa coupling' $G_{\rm W}$ to the Higgs field, similar to the lattice $U(1)_V \times U(1)_A$ model used by Luuk. However, presenting this was beyond the scope of the paper and it was also questionable whether decoupling doublers is possible by increasing $G_{\rm w}$. 

Results on dynamical breaking of chiral symmetry in the Hamiltonian formulation were published in 1980 also by Greensite and Primack using `naive fermions' \cite{Greensite:1980hy}\footnote{Practically simultaneously with \cite{Smit:1980nf} and with nearly the same title (!).}, and by Drell, Weinstein, Yankielowicz and Svetitsky \cite{Svetitsky:1980pa,Weinstein:1980jk}. Later that year Blairon, Brout, Englert and Greensite obtained similar results in the Euclidean formulation \cite{Blairon:1980pk}.

After the Santa Barbara conference I stayed a month with Luuk in Menlo Park near Stanford, writing together our `anomaly paper' which was submitted in October 1980 \cite{Karsten:1980wd}\footnote{Errata in \cite{Groot:1983ng}.}; in short:\\
- exact symmetry on the lattice $\Rightarrow$ no anomaly in the continuum limit;\\
- chiral anomaly $\Rightarrow$ need to break chiral symmetry initially on the lattice;\\
- in QCD, Wilson-fermions produce the flavour-$U(1)$ anomaly. 

During my stay Luuk told me that he had been studying topology aiming at formulating a no-go theorem for accommodating only left-handed (or right-handed) massless fermions (Weyl fermions\footnote{The familiar neutrinos are left handed Weyl fermions.})
on a Euclidean lattice. I was surprised and expressed not much interest.\footnote{At that time my understanding of the relevance of topology stopped at the fact in one dimension momentum space had the topology of a circle and the periodic $P_j(k)$ had to approach an even number of zeros with positive and negative unit slope, hence an equal number of positive and negative unit $Q_5$. In hind sight, being able to make statements for a generalised allowed implementation of $P_j(k)$ on different lattice types has been very important.}

The project became viable and easy to understand after unearthing the Poincar\'{e}-Hopf theorem. The theorem {\em equates} the \emph{index} of a vector field on a manifold with the \emph{Euler characteristic} $\ch_{\rm E}$ of the manifold. The typical example for the lattice is momentum space with topology of the hypertorus $T^4$, Euler number zero, and the vector field $P_\mu(k)$ has indices equal to $Q_5$ and index $\sum Q_5=\ch_{\rm E}=0$. 

November 1980 Luuk returned to Amsterdam. He gave talks on fermion doubling at various institutes in Europe. January 1981 he discussed with Holger Nielsen in Copenhagen. His no-go article was submitted in May that year \cite{Karsten:1981gd}. 
The famous no-go papers by Nielsen and Ninomiya \cite{Nielsen:1980rz}, \cite{Nielsen:1981xu} and \cite{Nielsen:1981hk} were submitted respectively November 1980 (final version January 1981), February 1981 (final version June 1981) and June 1981. 

Topology had become an important subject for `the lattice' also because of a resolution of the $U(1)$ problem in terms of the \emph{topological susceptibility} $\ch_{\rm top}$ by Witten \cite{Witten:1979vv} and Veneziano \cite{Veneziano:1979ec}. For three flavors it can be expressed in the form 
\[ 
m^2_{\eta^\prime} = \frac{6}{f^2_\pi}\,\ch_{\rm top}+ \half\,\left(m^2_{\eta} + m^2_{\pi^0}\right)\,,\quad
\ch_{\rm top}=\intx\, \langle q(x)\,q(0)\rangle_{\rm pure\, GT}\,.
\]
The \emph{topological charge density} $q(x)$ is the $U(1)$ flavor anomaly operator. 
The phenomenological value of the susceptibility is $\chtop \simeq (182\,\mbox{MeV})^4$. 
In 1981 a preliminary Monte Carlo estimate of $\chtop$ was presented by Di Vecchia, Fabricius, Rossi and Veneziano \cite{DiVecchia:1981aev} and several of its uncertainties where also investigated. Practically simultaneously L\"{u}scher submitted a paper on topology of lattice gauge fields \cite{Luscher:1981zq}. He formulated a smoothness-type condition on the lattice gluonic field strength which allowed defining a $q(x)$ giving topological charge $\Qtop=\intx\,q(x)$ that takes integer values. 
These papers initiated a whole new line of research on topology, instantons and the likes in lattice gauge theory.

An effective action at strong coupling QCD with staggered fermions as well as Wilson fermions in the Euclidean formulation was presented in 1981, together with Kawamoto \cite{Kawamoto:1981hw}. With Wilson's fermions there was again a critical $\Mc(g,r)$ and at $M=\Mc$  the effective action was chiral invariant to second order in the meson fields, but not in higher order as exemplified by the pion scattering amplitude. 
A similar effective action with `naive fermions' for $U(N)$ gauge theory was derived by Kluberg-Stern, Morel, Napoly and Petersson \cite{Kluberg-Stern:1981zya}, and by Alessandrini, Hakim and Krzywicki \cite{Alessandrini:1981gx}. 
Jaap Hoek joined in a companion paper to \cite{Kawamoto:1981hw} concentrating on baryons \cite{Hoek:1981uv}. 

Jaap had been developing computer algebra for deriving the effective hadron action for QCD in a strong coupling expansion with Wilson fermions \cite{Hoek:1982nw,Hoek:1983rx,Hoek:1985sr}. The resulting programs turned out to take too much computer resources to reach high orders in the expansion, unlike, for instance the impressive results obtained by M\"{u}nster in pure gauge theory \cite{Munster:1980iv,Munster:1981es,Munster:1982kg}.
After defending his PhD thesis \cite{Hoek:1983it} in October 1983, Jaap moved to the Rutherford Appleton Laboratory near Oxford. 

Around 1982 {\em mean field}  calculations by Greensite, Hansson, Hari Dass and Lauwers were also done at the National HEP Institute (Nikhef) in Amsterdam \cite{Greensite:1981ev,HariDass:1981fb}. Results using Monte Carlo methods were nicely explained and put in context by Kogut \cite{Kogut:1982ds}, and by Creutz, Jacobs and Rebbi \cite{Creutz:1983ev}. 

Returning to the phenomenon of fermion doubling, further insight came from:\\
- weak coupling calculations and definition of transfer matrix with staggered fermions by Sharatchandra, Thun and Weisz \cite{Sharatchandra:1981si};\\
- work on DWY fermions by Rabin \cite{Rabin:1981nm};\\
- relevance of algebraic geometry to lattice fermions by Rabin \cite{Rabin:1981qj};\\ 
- Dirac-K\"{a}hler fermions by Becher and Joos \cite{Becher:1981cb,Becher:1982ud};\\
- staggered fermion flavours and coupling to gauge fields by Gliozzi \cite{Gliozzi:1982ib};\\
- staggered fermion flavours by Kluberg-Stern, Morel, Napoly, Petersson \cite{Kluberg-Stern:1983lmr};\\
- order $1/g^2$ corrections with staggered fermions by Kluberg-Stern \emph{et.\ al.} \cite{Jolicoeur:1983tz};\\
- reduced staggered fermion calculations, transfer matrix,  with van den Doel\cite{vandenDoel:1983mf}.\\ 
How to use the insight of Ginsparg and Wilson on remnant chiral symmetry \cite{Ginsparg:1981bj} was a mystery to me.
 
Cees van den Doel (undergraduate student) drew my attention to papers by Banks \& Kaplunovsky \cite{Banks:1981ez} and Banks \& Zaks \cite{Banks:1981nn} which addressed the possibility of massless composite fermions in models with a real gauge group. We found that at strong but not infinite coupling the composite fermions come out massive by dynamical symmetry breaking \cite{vandenDoel:1983mf}. An important aspect of the paper is its spin and flavor interpretation derived from the symmetry group of Euclidean \emph{gauged} staggered fermions, including a reduction to two flavors and a calculational method in momentum space.\footnote{Staggered fermion fields at a single lattice point may have no Dirac and flavor indices --- these are somewhat spread out over the lattice. 
Their  identification has been done in differing ways and has sometimes influenced the way gauge fields are introduced.}

September 1983, while attending again a Carg\`{e}se Summer Institute at which Wilson lectured \cite{Wilson:1984vkc}, I mentioned to him my lattice gauge theory work in 1972-1973 at UCLA. His generous response was: \emph{Ask your PhD advisor to send me a description of what you did and I shall mention it in future reviews}. So it was done.\footnote{I had kept all my notes with detailed derivations and made a summary of these. Finkelstein (Bob) sent a letter to Wilson including the summary.}

Also in 1983, Maarten F.L.\ Golterman from Utrecht University became PhD student with me in Amsterdam. We continued deriving  results for staggered fermions: 
perturbative calculations \cite{Golterman:1984ds}, more on the flavor interpretation \cite{Golterman:1984cy} and baryon operators \cite{Golterman:1984dn}. 
As sole author Maarten presented meson operators \cite{Golterman:1985dz} and representations of the staggered fermion symmetry group \cite{Golterman:1986jf}. 
September 1986 he defended his PhD thesis \emph{Staggered Fermions} and soon after left for UC, Davies.

A comparison of various spin and flavor interpretations of staggered fermions was made by Daniel and Kieu \cite{Daniel:1986zm}.  Kilcup and Sharpe developed methods for evaluation of matrix elements of the effective electro-weak Hamiltonian \cite{Kilcup:1986dg}. Anomalies continued to stimulate perturbative checks such as a calculation of the non-Abelian anomaly with staggered and Wilson fermions, by Coste, Korthals Altes and Napoly \cite{Coste:1986qr}.

Yearly \emph{Lattice Conferences} started taking place. These were very important to the community, in which speakers felt free exposing unpublished work and ideas. Inspiring papers using Monte-Carlo methods were being published and I wished similar computational work was done in the Netherlands. The arrival in 1984 of the CDC Cyber 205 supercomputer in Amsterdam was a good occasion which led to taking part in larger collaborations:\\
-  computing hadron propagators with K\"{o}nig, M\"{u}tter, Schilling (Wupperthal)\cite{Konig:1985pw};\\
- with Bowler \emph{et.\ al.}\footnote{K.C.\ Bowler (Edinburgh), F.\ Gutbrod (DESY), P.\ Hasenfratz (Bern), U.\ Heller (CERN), F.\ Karsch (Urbana-Champaign), R.D.\ Kenway (Edinburgh), I.\ Montvay (DESY), G.S.\ Pawley (Edinburgh), J.\ Smit (Amsterdam), D.J.\ Wallace (Edinburgh).}  
on the $\bt$-function and the quark-antiquark potential \cite{Bowler:1985ev}. 
Peter Hasenfratz sent me gauge-field configurations, announcing \emph{``They are beautiful !''} .\\[2mm]
Taking part in these `computer experiments' gave an excitement similar to applying Regge pole theory to hadron scattering experiments. The  potential turned out to be surprisingly linear after subtraction of the Coulomb type term $-\pi/(12 r)$ calculated by L\"{u}scher for the fluctuating string model \cite{Luscher:1980ac}. The {\em roughening transition} in the string tension was studied further by Kogut and Sinclair \cite{Kogut:1981yq}. 

In 1986 remarkable work appeared by L\"{u}scher in which finite-size effects were exploited for the computation of scattering phase shifts \cite{Luscher:1986pf} and related work on properties of unstable particles \cite{Luscher:1991cf}. 

The computer became a great help, together with Jeroen Vink (new PhD student),
for research on lattice fermion aspects of the $U(1)$ problem. A lattice derivation of the Witten-Veneziano relation led to a `fermionic definition' of topological charge \cite{Smit:1987wa,Smit:1987zh}. For example for Wilson fermions
\[ 
\overline{Q}(U) = \Tr\,\kp_P\,m\,\gm_5\,\left[D(U) + m + \Mc - W(U)\right]^{-1}\,, 
\]
where $\kp_P$ is a finite renormalization. In continuum field theory the $\Mc-W$ term is absent, $\kp_P=1$, and evaluating the trace operation $\Tr$ in terms of eigenvalues 
of the Dirac operator $D(U)$ only the $n_\pm$ zero modes of chirality $\pm$ contribute, such that the right-hand-side is the index of the Dirac operator. Then by the Atiya-Singer index theorem $\overline{Q}(U)$ is the topological charge of gauge field configuration $U$. We studied remnants of the index theorem on the lattice \cite{Smit:1986fn}. Jeroen computed the susceptibility $\overline{\chtop}$ in QCD and compared it with results obtained with the cooling method by Hoek, Teper and Waterhouse \cite{Hoek:1986nd}. The cooling method was introduced by Berg in \cite{Berg:1981nw}. Edinburgh gauge-field configurations \cite{Bowler:1985ev} were used and the necessary staggered fermion propagators were computed with the pseudo-fermion algorithm proposed by Fucito, Marinari, Parisi and Rebbi \cite{Fucito:1980fh}. 

In 1984 Swift published a proposal for a lattice regularization of the Standard Model (SM) using Wilson fermions \cite{Swift:1984dg}. It was identical to the one used for deriving currents in the `spin wave paper' \cite{Smit:1980nf} (reviewed in \cite{Smit:1985nu}). The model was obtained by replacing in fermion mass terms $\psb (M -W )\ps$, $M$ by the SM Yukawa couplings to the Higgs field with coupling constants, say $G$, and replacing the $r$-parameter in $W$ by similar terms with one coupling constant, say $G_{\rm W}$. 
The question is then whether the doublers can be removed by a suitable choice of $G_{\rm W}$ and the $G$s. 
For a start in answering this question the gauge fields may be left out, provided the Higgs field is given its usual vacuum expectation value.\footnote{According to D'Hoker and Fahri, possible anomalies may be carried by the Higgs field \cite{DHoker:1984mif}.} 

Shortly after Swift's paper Hands and Carpenter concluded from a study of a $U(1)$ staggered fermion model that the doublers cannot be removed, at best their masses can be raised somewhat above the wanted fermion mass \cite{Hands:1985ys}. 

Relevant in this context is that L\"{u}scher and Weisz showed in a series of papers starting with \cite{Luscher:1987ay} that it is highly unlikely that the renormalized Higgs selfcoupling in the Standard Model can be strong \cite{Luscher:1988gc}.

A lively collaboration developed with Jersak's group in Aachen, including Bock, De, Jansen and Neuhaus,  who studied Yukawa models numerically at the Supercomputing Centre in J\"{u}lich. The collaboration started out informally: every four weeks or so I made trip to J\"{u}lich to discuss new results. Later on Wolfgang Bock became a post doc in Amsterdam. The phase diagram of the Wilson-Yukawa model turned out to be quite complicated, with {\em two disjoint}  paramagnetic (symmetric) phases at weak and strong Yukawa couplings, respectively called PMW and PMS, and {\em one} ferromagnetic (broken symmetry) phase with a crossover separating weak and strong coupling regions called FM(W) and FM(S) \cite{Bock:1990tv}.\footnote{Choosing for simplicity equal values of the $G$s the resulting Wilson-Yukawa model is $SU(2)_{\rm L}\times SU(2)_{\rm R}$ invariant. 
The $SU(2)_{\rm L}$ factor corresponds to the $SU(2)$ in the Standard Model group $U(1)\times SU(2) \times SU(3)$. }
The phase boundary line between broken and symmetric phases had disjoint weak and strong coupling branches. Conventional physics had to be found near the weak coupling branch but it included fermion doublers \cite{Bock:1991kn}. Near the strong coupling branch the doublers could be removed but in the process the remaining fermion particles turned out to become {\em neutral} under $SU(2)_{\rm L}$, chargeless fermion-Higgs bound states! It took a few years and with help of Wolfgang Bock (at that time postdoc in Amsterdam) and Asit De, and with Maarten Golterman (then at UCLA) and Don Petcher (St. Louis) to suggest this interpretation as the most probable outcome \cite{Bock:1991bu,Golterman:1991re}.\footnote{This was not without controversy: \cite{Bock:1991bu} was submitted 1 November 1991,  accepted 1 September 1992.} The `standard lattice electroweak theory' was also studied by Aoki, Lee, Shigemitsu and Shrock, e.g.\ in \cite{Aoki:1990yb,Aoki:1991es}.

After all, the option proposed by Montvay of interpreting doublers as mirror fermions appeared to fare better (give larger mirror masses) \cite{Frick:1992ef}.

Gradually the quest for a gauge-invariant lattice formulation of {\em chiral gauge theory} with non-trivial anomaly cancelation became stronger. 
In 1989 Borrelli, Maiani, Sisto, Rossi and Testa had proposed using Wilson fermions with gauge fixing and suitable renormalization to recover gauge invariance in the continuum limit \cite{Borrelli:1989kw}. Avoiding gauge fixing would facilitate much more easily numerical simulations. A proposal gauging staggered fermion flavors did not work out well \cite{Bock:1993qm}.\footnote{The physical interpretation of staggered fermions requires smoothly varying Bose fields  whereas gauge transformations may behave wildly, which reminds me of Giorgio Parisi's comment on the proposal at Lattice 89: \emph{This is very dangerous !}}
Banks and Dabholkar analysed doubler decoupling in the Standard Model more generally and argued that the Wilson-Yukawa approach fails \cite{Banks:1992af} but an approach as studied in \cite{Bock:1993qm} might not.

Game changer David Kaplan alerted me at lunch during Lattice 1992 about his paper 
\emph{A Method for simulating chiral fermions on the lattice} \cite{Kaplan:1992bt,Kaplan:1992sg}. It opened the door to new progress and led also to an unexpected appreciation of the fundamental work by Ginsparg and Wilson \cite{Ginsparg:1981bj}, for example in L\"{u}scher's article \cite{Luscher:1998pqa}. 

These recollections spanned about the first 20 years of lattice QCD and were influenced much by chiral symmetry with its anomalies and fermion doubling.   
Cross-fertilizations with `theoretical theory' (e.g.\ on confinement) or algorithms (e.g.\ multigrid) were barely or not mentioned.
Staggered fermion flavors have been re-named {\em `tastes'} with a prescription for compensating their $N_{\rm taste}$-fold multiplicity by an algorithm that corresponds to multiplying fermion loops of a given flavor by $1/N_{\rm taste}$ \cite{Hamber:1983kx}. This caused some controversy, nicely reviewed by Andreas Kronfeld \cite{Kronfeld:2007ek}. For QCD, Wilson fermions are free of controversy (to my knowledge), 
albeit requiring more computer resources.
I believe that the continuum limit of Wilson's original formulation gives a {\em definition} of Euclidean QCD.

As exemplified by {\em Lattice 2024}, the next thirty years continued showing new theoretical developments and progress in computation. 
Lattice QCD has become also indispensable for interpreting experimental results in the electroweak sector of the Standard Model.\\

Added Note. I had been unable to access proceedings of the conference mentioned in footnote 10, except very recently on a website of the U.S. Department of Energy (www.osti.gov).
The proceedings of the {\em Marseille Colloquium on Lagrangian Field Theory}, June 24-28, 1974, contain contributions on lattice gauge theory by Kenneth Wilson \cite{Wilson:1974evp} and by Chris Korthals Altes \cite{Korthals-Altes:1974xaz}; Jean-Michel Drouffe summarized work with Balian and Itzykson in the form of abstracts of publications. Chris heard September 1973 about lattice gauge theory from Gerard 't Hooft who attended a seminar at CERN by Wilson in August that year.

\renewcommand{\baselinestretch}{0.5}
\small\small
\setlength{\baselineskip}{2pt}

\bibliography{lit}

\providecommand{\href}[2]{#2}\begingroup\raggedright\begin{thebibliography}{100}

\bibitem{Veltman:1968ki}
M.J.G.~Veltman, \emph{{Perturbation theory of massive Yang-Mills fields}},
  \href{https://doi.org/10.1016/0550-3213(68)90197-1}{\emph{Nucl. Phys. B}
  {\bfseries 7} (1968) 637}.

\bibitem{Weinberg:1967tq}
S.~Weinberg, \emph{{A Model of Leptons}},
  \href{https://doi.org/10.1103/PhysRevLett.19.1264}{\emph{Phys. Rev. Lett.}
  {\bfseries 19} (1967) 1264}.

\bibitem{Wilson:1969zs}
K.G.~Wilson, \emph{{Nonlagrangian models of current algebra}},
  \href{https://doi.org/10.1103/PhysRev.179.1499}{\emph{Phys. Rev.} {\bfseries
  179} (1969) 1499}.

\bibitem{Wilson:1970pq}
K.G.~Wilson, \emph{{Operator Product Expansions and Anomalous Dimensions in the
  Thirring Model}}, \href{https://doi.org/10.1103/PhysRevD.2.1473}{\emph{Phys.
  Rev. D} {\bfseries 2} (170) 1473}.

\bibitem{Wilson:1970ag}
K.G.~Wilson, \emph{{The Renormalization Group and Strong Interactions}},
  \href{https://doi.org/10.1103/PhysRevD.3.1818}{\emph{Phys. Rev. D} {\bfseries
  3} (1971) 1818}.

\bibitem{Wilson:1971bg}
K.G.~Wilson, \emph{{Renormalization group and critical phenomena I.
  Renormalization group and the Kadanoff scaling picture}},
  \href{https://doi.org/10.1103/PhysRevB.4.3174}{\emph{Phys. Rev. B} {\bfseries
  4} (1971) 3174}.

\bibitem{Wilson:1971dh}
K.G.~Wilson, \emph{{Renormalization group and critical phenomena II. Phase
  space cell analysis of critical behavior}},
  \href{https://doi.org/10.1103/PhysRevB.4.3184}{\emph{Phys. Rev. B} {\bfseries
  4} (1971) 3184}.

\bibitem{Smit:1971zz}
J.~Smit, \emph{{Conformal Invariance and Indefinite Metric}},
  \href{https://doi.org/10.13140/RG.2.2.24907.91680}{\emph{ITFA-71-2200} (1971)
  }.

\bibitem{Wiegel:1977kp}
F.W.~Wiegel, \emph{{The Interacting Bose Fluid: Path Integral Representations
  and Renormalization Group Approach}}, {\emph{NATO Sci. Ser. B} {\bfseries 34}
  (1978) 419}.

\bibitem{Boulware:1970zc}
D.G.~Boulware, \emph{{Renormalizeability of massive non-abelian gauge fields -
  a functional integral approach}},
  \href{https://doi.org/10.1016/0003-4916(70)90008-4}{\emph{Annals Phys.}
  {\bfseries 56} (1970) 140}.

\bibitem{Rzewuski}
J.~Rzewuski, \emph{{Field Theory}}, Iliffe Books, London (1969).

\bibitem{Adler:1969gk}
S.L.~Adler, \emph{{Axial vector vertex in spinor electrodynamics}},
  \href{https://doi.org/10.1103/PhysRev.177.2426}{\emph{Phys. Rev.} {\bfseries
  177} (1969) 2426}.

\bibitem{Bell:1969ts}
J.S.~Bell and R.~Jackiw, \emph{{A PCAC puzzle: $\pi^0 \to \gamma \gamma$ in the
  $\sigma$ model}}, \href{https://doi.org/10.1007/BF02823296}{\emph{Nuovo Cim.
  A} {\bfseries 60} (1969) 47}.

\bibitem{Bouchiat:1972iq}
C.~Bouchiat, J.~Iliopoulos and P.~Meyer, \emph{{An Anomaly Free Version of
  Weinberg's Model}},
  \href{https://doi.org/10.1016/0370-2693(72)90532-1}{\emph{Phys. Lett. B}
  {\bfseries 38} (1972) 519}.

\bibitem{Gross:1972pv}
D.J.~Gross and R.~Jackiw, \emph{{Effect of anomalies on quasirenormalizable
  theories}}, \href{https://doi.org/10.1103/PhysRevD.6.477}{\emph{Phys. Rev. D}
  {\bfseries 6} (1972) 477}.

\bibitem{Adler:1970Brandeis}
S.L.~Adler, \emph{{Perturbation Theory Anomalies}}, Brandeis University Summer
  Institute in Theoretical Physics, MIT press (1970).

\bibitem{Schwinger:1969vz}
J.~Schwinger, \emph{{Particles and Sources}}, Addison-Wesley (1969).

\bibitem{Schwinger:1970xc}
J.~Schwinger, \emph{{Particles, Sources, and Fields. Volume I}}, Addison-Wesley
  (1970).

\bibitem{Schwinger:1973rv}
J.S.~Schwinger, \emph{{Particles, Sources and Fields. Volume II}},
  Addison-Wesley (1973).

\bibitem{Finkelstein:1974xm}
R.J.~Finkelstein and J.~Smit, \emph{{Massive gauge field in source theory.
  2.}}, \href{https://doi.org/10.1016/0003-4916(74)90401-1}{\emph{Annals Phys.}
  {\bfseries 88} (1974) 157}.

\bibitem{Cornwall:1973ts}
J.M.~Cornwall and R.E.~Norton, \emph{{Spontaneous Symmetry Breaking Without
  Scalar Mesons}}, \href{https://doi.org/10.1103/PhysRevD.8.3338}{\emph{Phys.
  Rev. D} {\bfseries 8} (1973) 3338}.

\bibitem{Smit:1974je}
J.~Smit, \emph{{On the Possibility That Massless Yang-Mills Fields Generate
  Massive Vector Particles}},
  \href{https://doi.org/10.1103/PhysRevD.10.2473}{\emph{Phys. Rev. D}
  {\bfseries 10} (1974) 2473}.

\bibitem{Wilson:1973jj}
K.G.~Wilson and J.B.~Kogut, \emph{{The Renormalization group and the epsilon
  expansion}}, \href{https://doi.org/10.1016/0370-1573(74)90023-4}{\emph{Phys.
  Rept.} {\bfseries 12} (1974) 75}.

\bibitem{Wilson:1974sk}
K.G.~Wilson, \emph{{Confinement of Quarks}},
  \href{https://doi.org/10.1103/PhysRevD.10.2445}{\emph{Phys. Rev. D}
  {\bfseries 10} (1974) 2445}.

\bibitem{Drell:1976mj}
S.D.~Drell, M.~Weinstein and S.~Yankielowicz, \emph{{Strong Coupling Field
  Theories II, Fermions and Gauge Fields on a Lattice}},
  \href{https://doi.org/10.1103/PhysRevD.14.1627}{\emph{Phys. Rev. D}
  {\bfseries 14} (1976) 1627}.

\bibitem{Wilson:1975id}
K.G.~Wilson, \emph{{Quarks and Strings on a Lattice}},  in \emph{{13th
  International School of Subnuclear Physics: New Phenomena in Subnuclear
  Physics}}, 11, 1975.

\bibitem{Kogut:1974ag}
J.B.~Kogut and L.~Susskind, \emph{{Hamiltonian Formulation of Wilson's Lattice
  Gauge Theories}}, \href{https://doi.org/10.1103/PhysRevD.11.395}{\emph{Phys.
  Rev. D} {\bfseries 11} (1975) 395}.

\bibitem{Susskind:1976jm}
L.~Susskind, \emph{{Lattice Fermions}},
  \href{https://doi.org/10.1103/PhysRevD.16.3031}{\emph{Phys. Rev. D}
  {\bfseries 16} (1977) 3031}.

\bibitem{Balian:1974xw}
R.~Balian, J.M.~Drouffe and C.~Itzykson, \emph{{Gauge Fields on a Lattice. III.
  Strong Coupling Expansions and Transition Points}},
  \href{https://doi.org/10.1103/PhysRevD.11.2104}{\emph{Phys. Rev. D}
  {\bfseries 11} (1975) 2104}.

\bibitem{Banks:1976ia}
{\scshape Cornell-Oxford-Tel Aviv-Yeshiva} collaboration, \emph{{Strong
  Coupling Calculations of the Hadron Spectrum of Quantum Chromodynamics}},
  \href{https://doi.org/10.1103/PhysRevD.15.1111}{\emph{Phys. Rev. D}
  {\bfseries 15} (1977) 1111}.

\bibitem{Sharatchandra:1976af}
H.S.~Sharatchandra, \emph{{The Continuum Limit of Lattice Gauge Theories in the
  Context of Renormalized Perturbation Theory}},
  \href{https://doi.org/10.1103/PhysRevD.18.2042}{\emph{Phys. Rev. D}
  {\bfseries 18} (1978) 2042}.

\bibitem{Baaquie:1977hz}
B.E.~Baaquie, \emph{{Gauge Fixing and Mass Renormalization in the Lattice Gauge
  Theory}}, \href{https://doi.org/10.1103/PhysRevD.16.2612}{\emph{Phys. Rev. D}
  {\bfseries 16} (1977) 2612}.

\bibitem{Creutz:1976ch}
M.~Creutz, \emph{{Gauge Fixing, the Transfer Matrix, and Confinement on a
  Lattice}}, \href{https://doi.org/10.1103/PhysRevD.15.1128}{\emph{Phys. Rev.
  D} {\bfseries 15} (1977) 1128}.

\bibitem{Luscher:1976ms}
M.~Luscher, \emph{{Construction of a Selfadjoint, Strictly Positive Transfer
  Matrix for Euclidean Lattice Gauge Theories}},
  \href{https://doi.org/10.1007/BF01614090}{\emph{Commun. Math. Phys.}
  {\bfseries 54} (1977) 283}.

\bibitem{Karsten:1978nb}
L.H.~Karsten and J.~Smit, \emph{{Axial Symmetry in Lattice Theories}},
  \href{https://doi.org/10.1016/0550-3213(78)90385-1}{\emph{Nucl. Phys. B}
  {\bfseries 144} (1978) 536}.

\bibitem{Karsten:1979wh}
L.H.~Karsten and J.~Smit, \emph{{The Vacuum Polarization With {SLAC} Lattice
  Fermions}}, \href{https://doi.org/10.1016/0370-2693(79)90786-X}{\emph{Phys.
  Lett. B} {\bfseries 85} (1979) 100}.

\bibitem{Shigemitsu:1978br}
J.~Shigemitsu, \emph{{Spectrum Calculations in Lattice Gauge Theory Using
  Wilson's Fermion Method}},
  \href{https://doi.org/10.1103/PhysRevD.18.1709}{\emph{Phys. Rev. D}
  {\bfseries 18} (1978) 1709}.

\bibitem{Greensite:1979ha}
J.P.~Greensite, \emph{{Large Scale Vacuum Structure and New Calculational
  Techniques in Lattice SU($N$) Gauge Theory}},
  \href{https://doi.org/10.1016/0550-3213(80)90494-0}{\emph{Nucl. Phys. B}
  {\bfseries 166} (1980) 113}.

\bibitem{Savit:1979ny}
R.~Savit, \emph{{Duality in Field Theory and Statistical Systems}},
  \href{https://doi.org/10.1103/RevModPhys.52.453}{\emph{Rev. Mod. Phys.}
  {\bfseries 52} (1980) 453}.

\bibitem{Banks:1977cc}
T.~Banks, R.~Myerson and J.B.~Kogut, \emph{{Phase Transitions in Abelian
  Lattice Gauge Theories}},
  \href{https://doi.org/10.1016/0550-3213(77)90129-8}{\emph{Nucl. Phys. B}
  {\bfseries 129} (1977) 493}.

\bibitem{Fradkin:1978dv}
E.H.~Fradkin and S.H.~Shenker, \emph{{Phase Diagrams of Lattice Gauge Theories
  with Higgs Fields}},
  \href{https://doi.org/10.1103/PhysRevD.19.3682}{\emph{Phys. Rev. D}
  {\bfseries 19} (1979) 3682}.

\bibitem{Mack:1978kr}
G.~Mack and V.B.~Petkova, \emph{{Sufficient Condition for Confinement of Static
  Quarks by a Vortex Condensation Mechanism}},
  \href{https://doi.org/10.1016/0003-4916(80)90121-9}{\emph{Annals Phys.}
  {\bfseries 125} (1980) 117}.

\bibitem{Munster:1979pg}
G.~Munster, \emph{{On the Characterization of the Higgs Phase in Lattice Gauge
  Theories}}, \href{https://doi.org/10.1007/BF01588845}{\emph{Z. Phys. C}
  {\bfseries 6} (1980) 175}.

\bibitem{Wilson:1979wp}
K.G.~Wilson, \emph{{Monte Carlo Calculations for the Lattice Gauge Theory}},
  \href{https://doi.org/10.1007/978-1-4684-7571-5_20}{\emph{NATO Sci. Ser. B}
  {\bfseries 59} (1980) 363}.

\bibitem{Kogut:1979vg}
J.B.~Kogut, R.B.~Pearson and J.~Shigemitsu, \emph{{The QCD beta Function at
  Intermediate and Strong Coupling}},
  \href{https://doi.org/10.1103/PhysRevLett.43.484}{\emph{Phys. Rev. Lett.}
  {\bfseries 43} (1979) 484}.

\bibitem{Creutz:1979zg}
M.~Creutz, L.~Jacobs and C.~Rebbi, \emph{{Monte Carlo Study of Abelian Lattice
  Gauge Theories}}, \href{https://doi.org/10.1103/PhysRevD.20.1915}{\emph{Phys.
  Rev. D} {\bfseries 20} (1979) 1915}.

\bibitem{Creutz:1979cu}
M.~Creutz, \emph{{Solving Quantized SU(2) Gauge Theory}}, {\emph{Print-79-0919
  (BNL)} (1979) }.

\bibitem{Creutz:1980zw}
M.~Creutz, \emph{{Monte Carlo Study of Quantized SU(2) Gauge Theory}},
  \href{https://doi.org/10.1103/PhysRevD.21.2308}{\emph{Phys. Rev. D}
  {\bfseries 21} (1980) 2308}.

\bibitem{Creutz:1980wj}
M.~Creutz, \emph{{Asymptotic Freedom Scales}},
  \href{https://doi.org/10.1103/PhysRevLett.45.313}{\emph{Phys. Rev. Lett.}
  {\bfseries 45} (1980) 313}.

\bibitem{Karsten:1979qs}
L.H.~Karsten, \emph{{The Lattice Fermion Problem and Weak Coupling Perturbation
  Theory}}, {\emph{NATO Sci. Ser. B} {\bfseries 55} (1980) 235}.

\bibitem{Banks:1979yr}
T.~Banks and A.~Casher, \emph{{Chiral Symmetry Breaking in Confining
  Theories}}, \href{https://doi.org/10.1016/0550-3213(80)90255-2}{\emph{Nucl.
  Phys. B} {\bfseries 169} (1980) 103}.

\bibitem{Kawamoto:1980fd}
N.~Kawamoto, \emph{{Towards the Phase Structure of Euclidean Lattice Gauge
  Theories with Fermions}},
  \href{https://doi.org/10.1016/0550-3213(81)90450-8}{\emph{Nucl. Phys. B}
  {\bfseries 190} (1981) 617}.

\bibitem{Smit:1980nf}
J.~Smit, \emph{{Chiral Symmetry Breaking in QCD: Mesons as Spin Waves}},
  \href{https://doi.org/10.1016/0550-3213(80)90056-5}{\emph{Nucl. Phys. B}
  {\bfseries 175} (1980) 307}.

\bibitem{Greensite:1980hy}
J.~Greensite and J.~Primack, \emph{{Pions as Spin Waves: Chiral Symmetry
  Breaking in Lattice Gauge Theory}},
  \href{https://doi.org/10.1016/0550-3213(81)90160-7}{\emph{Nucl. Phys. B}
  {\bfseries 180} (1981) 170}.

\bibitem{Svetitsky:1980pa}
B.~Svetitsky, S.D.~Drell, H.R.~Quinn and M.~Weinstein, \emph{{Dynamical
  Breaking of Chiral Symmetry in Lattice Gauge Theories}},
  \href{https://doi.org/10.1103/PhysRevD.22.490}{\emph{Phys. Rev. D} {\bfseries
  22} (1980) 490}.

\bibitem{Weinstein:1980jk}
M.~Weinstein, S.D.~Drell, H.R.~Quinn and B.~Svetitsky, \emph{{Approximate
  Dynamical Symmetry in Lattice Quantum Chromodynamics}},
  \href{https://doi.org/10.1103/PhysRevD.22.1190}{\emph{Phys. Rev. D}
  {\bfseries 22} (1980) 1190}.

\bibitem{Blairon:1980pk}
J.M.~Blairon, R.~Brout, F.~Englert and J.~Greensite, \emph{{Chiral Symmetry
  Breaking in the Action Formulation of Lattice Gauge Theory}},
  \href{https://doi.org/10.1016/0550-3213(81)90061-4}{\emph{Nucl. Phys. B}
  {\bfseries 180} (1981) 439}.

\bibitem{Karsten:1980wd}
L.H.~Karsten and J.~Smit, \emph{{Lattice Fermions: Species Doubling, Chiral
  Invariance, and the Triangle Anomaly}},
  \href{https://doi.org/10.1016/0550-3213(81)90549-6}{\emph{Nucl. Phys. B}
  {\bfseries 183} (1981) 103}.

\bibitem{Groot:1983ng}
R.~Groot, J.~Hoek and J.~Smit, \emph{{Normalization of Currents in Lattice
  QCD}}, \href{https://doi.org/10.1016/0550-3213(84)90018-X}{\emph{Nucl. Phys.
  B} {\bfseries 237} (1984) 111}.

\bibitem{Karsten:1981gd}
L.H.~Karsten, \emph{{Lattice Fermions in Euclidean Space-time}},
  \href{https://doi.org/10.1016/0370-2693(81)90133-7}{\emph{Phys. Lett. B}
  {\bfseries 104} (1981) 315}.

\bibitem{Nielsen:1980rz}
H.B.~Nielsen and M.~Ninomiya, \emph{{Absence of Neutrinos on a Lattice. 1.
  Proof by Homotopy Theory}},
  \href{https://doi.org/10.1016/0550-3213(82)90011-6}{\emph{Nucl. Phys. B}
  {\bfseries 185} (1981) 20}.

\bibitem{Nielsen:1981xu}
H.B.~Nielsen and M.~Ninomiya, \emph{{Absence of Neutrinos on a Lattice. 2.
  Intuitive Topological Proof}},
  \href{https://doi.org/10.1016/0550-3213(81)90524-1}{\emph{Nucl. Phys. B}
  {\bfseries 193} (1981) 173}.

\bibitem{Nielsen:1981hk}
H.B.~Nielsen and M.~Ninomiya, \emph{{No Go Theorem for Regularizing Chiral
  Fermions}}, \href{https://doi.org/10.1016/0370-2693(81)91026-1}{\emph{Phys.
  Lett. B} {\bfseries 105} (1981) 219}.

\bibitem{Witten:1979vv}
E.~Witten, \emph{{Current Algebra Theorems for the U(1) Goldstone Boson}},
  \href{https://doi.org/10.1016/0550-3213(79)90031-2}{\emph{Nucl. Phys. B}
  {\bfseries 156} (1979) 269}.

\bibitem{Veneziano:1979ec}
G.~Veneziano, \emph{{U(1) Without Instantons}},
  \href{https://doi.org/10.1016/0550-3213(79)90332-8}{\emph{Nucl. Phys. B}
  {\bfseries 159} (1979) 213}.

\bibitem{DiVecchia:1981aev}
P.~Di~Vecchia, K.~Fabricius, G.C.~Rossi and G.~Veneziano, \emph{{Preliminary
  Evidence for U(1)-A Breaking in QCD from Lattice Calculations}},
  \href{https://doi.org/10.1016/0550-3213(81)90432-6}{\emph{Nucl. Phys. B}
  {\bfseries 192} (1981) 392}.

\bibitem{Luscher:1981zq}
M.~Luscher, \emph{{Topology of Lattice Gauge Fields}},
  \href{https://doi.org/10.1007/BF02029132}{\emph{Commun. Math. Phys.}
  {\bfseries 85} (1982) 39}.

\bibitem{Kawamoto:1981hw}
N.~Kawamoto and J.~Smit, \emph{{Effective Lagrangian and Dynamical Symmetry
  Breaking in Strongly Coupled Lattice QCD}},
  \href{https://doi.org/10.1016/0550-3213(81)90196-6}{\emph{Nucl. Phys. B}
  {\bfseries 192} (1981) 100}.

\bibitem{Kluberg-Stern:1981zya}
H.~Kluberg-Stern, A.~Morel, O.~Napoly and B.~Petersson, \emph{{Spontaneous
  Chiral Symmetry Breaking for a U(N) Gauge Theory on a Lattice}},
  \href{https://doi.org/10.1016/0550-3213(81)90445-4}{\emph{Nucl. Phys. B}
  {\bfseries 190} (1981) 504}.

\bibitem{Alessandrini:1981gx}
V.~Alessandrini, V.~Hakim and A.~Krzywicki, \emph{{Chiral Symmetry in the Mean
  Field Approach to Large $N$ Lattice {QCD}}},
  \href{https://doi.org/10.1016/0550-3213(82)90388-1}{\emph{Nucl. Phys. B}
  {\bfseries 205} (1982) 253}.

\bibitem{Hoek:1981uv}
J.~Hoek, N.~Kawamoto and J.~Smit, \emph{{Baryons in the Effective Lagrangian of
  Strongly Coupled Lattice {QCD}}},
  \href{https://doi.org/10.1016/0550-3213(82)90357-1}{\emph{Nucl. Phys. B}
  {\bfseries 199} (1982) 495}.

\bibitem{Hoek:1982nw}
J.~Hoek, \emph{{Strong Coupling Expansion of the Generating Functional for
  Gauge Systems on a Lattice With Arbitrary Sources}},
  \href{https://doi.org/10.1016/0021-9991(83)90126-2}{\emph{J. Comput. Phys.}
  {\bfseries 49} (1983) 265}.

\bibitem{Hoek:1983rx}
J.~Hoek, \emph{{Strong Coupling Expansion Of The SU(3) AND U(3) Effective
  Actions}}, \href{https://doi.org/10.1016/0021-9991(84)90117-7}{\emph{J.
  Comput. Phys.} {\bfseries 54} (1984) 245}.

\bibitem{Hoek:1985sr}
J.~Hoek and J.~Smit, \emph{{On the 1/$g^2$ Corrections to Hadron Masses}},
  \href{https://doi.org/10.1016/0550-3213(86)90031-3}{\emph{Nucl. Phys. B}
  {\bfseries 263} (1986) 129}.

\bibitem{Munster:1980iv}
G.~Munster, \emph{{High Temperature Expansions for the Free Energy of Vortices,
  Respectively the String Tension in Lattice Gauge Theories}},
  \href{https://doi.org/10.1016/0550-3213(81)90153-X}{\emph{Nucl. Phys. B}
  {\bfseries 180} (1981) 23}.

\bibitem{Munster:1981es}
G.~Munster, \emph{{Strong Coupling Expansions for the Mass Gap in Lattice Gauge
  Theories}}, \href{https://doi.org/10.1016/0550-3213(81)90570-8}{\emph{Nucl.
  Phys. B} {\bfseries 190} (1981) 439}.

\bibitem{Munster:1982kg}
G.~Munster, \emph{{Physical Strong Coupling Expansion Parameters and Glueball
  Mass Ratios}},
  \href{https://doi.org/10.1016/0370-2693(83)90201-0}{\emph{Phys. Lett. B}
  {\bfseries 121} (1983) 53}.

\bibitem{Hoek:1983it}
J.~Hoek, \emph{{Effective Action Calculation in Lattice QCD}}, Ph.D. thesis,
  Amsterdam U., 1983.

\bibitem{Greensite:1981ev}
J.~Greensite, T.H.~Hansson, N.D.~Hari~Dass and P.G.~Lauwers,
  \emph{{Calculations in the Weak and Crossover Regions of SU(2) Lattice Gauge
  Theory}}, \href{https://doi.org/10.1016/0370-2693(81)91021-2}{\emph{Phys.
  Lett. B} {\bfseries 105} (1981) 201}.

\bibitem{HariDass:1981fb}
N.D.~Hari~Dass and P.G.~Lauwers, \emph{{Some Approximate Calculations in SU(2)
  Lattice Mean Field Theory}},
  \href{https://doi.org/10.1016/0550-3213(82)90128-6}{\emph{Nucl. Phys. B}
  {\bfseries 210} (1982) 388}.

\bibitem{Kogut:1982ds}
J.B.~Kogut, \emph{{A Review of the Lattice Gauge Theory Approach to Quantum
  Chromodynamics}}, \href{https://doi.org/10.1103/RevModPhys.55.775}{\emph{Rev.
  Mod. Phys.} {\bfseries 55} (1983) 775}.

\bibitem{Creutz:1983ev}
M.~Creutz, L.~Jacobs and C.~Rebbi, \emph{{Monte Carlo Computations in Lattice
  Gauge Theories}},
  \href{https://doi.org/10.1016/0370-1573(83)90016-9}{\emph{Phys. Rept.}
  {\bfseries 95} (1983) 201}.

\bibitem{Sharatchandra:1981si}
H.S.~Sharatchandra, H.J.~Thun and P.~Weisz, \emph{{Susskind Fermions on a
  Euclidean Lattice}},
  \href{https://doi.org/10.1016/0550-3213(81)90200-5}{\emph{Nucl. Phys. B}
  {\bfseries 192} (1981) 205}.

\bibitem{Rabin:1981nm}
J.M.~Rabin, \emph{{Perturbation Theory for {SLAC} Lattice Fermions}},
  \href{https://doi.org/10.1103/PhysRevD.24.3218}{\emph{Phys. Rev. D}
  {\bfseries 24} (1981) 3218}.

\bibitem{Rabin:1981qj}
J.M.~Rabin, \emph{{Homology Theory of Lattice Fermion Doubling}},
  \href{https://doi.org/10.1016/0550-3213(82)90434-5}{\emph{Nucl. Phys. B}
  {\bfseries 201} (1982) 315}.

\bibitem{Becher:1981cb}
P.~Becher, \emph{{Dirac Fermions on the Lattice: A Local Approach Without
  Spectrum Degeneracy}},
  \href{https://doi.org/10.1016/0370-2693(81)90595-5}{\emph{Phys. Lett. B}
  {\bfseries 104} (1981) 221}.

\bibitem{Becher:1982ud}
P.~Becher and H.~Joos, \emph{{The Dirac-K\"{a}hler Equation and Fermions on the
  Lattice}}, \href{https://doi.org/10.1007/BF01614426}{\emph{Z. Phys. C}
  {\bfseries 15} (1982) 343}.

\bibitem{Gliozzi:1982ib}
F.~Gliozzi, \emph{{Spinor Algebra of the One Component Lattice Fermions}},
  \href{https://doi.org/10.1016/0550-3213(82)90199-7}{\emph{Nucl. Phys. B}
  {\bfseries 204} (1982) 419}.

\bibitem{Kluberg-Stern:1983lmr}
H.~Kluberg-Stern, A.~Morel, O.~Napoly and B.~Petersson, \emph{{Flavors of
  Lagrangian Susskind Fermions}},
  \href{https://doi.org/10.1016/0550-3213(83)90501-1}{\emph{Nucl. Phys. B}
  {\bfseries 220} (1983) 447}.

\bibitem{Jolicoeur:1983tz}
T.~Jolicoeur, H.~Kluberg-Stern, M.~Lev, A.~Morel and B.~Petersson, \emph{{The
  Strong Coupling Expansion of Lattice Gauge Theories With Susskind Fermions}},
  \href{https://doi.org/10.1016/0550-3213(84)90492-9}{\emph{Nucl. Phys. B}
  {\bfseries 235} (1984) 455}.

\bibitem{vandenDoel:1983mf}
C.~van~den Doel and J.~Smit, \emph{{Dynamical Symmetry Breaking in Two Flavor
  SU($N$) and SO($N$) Lattice Gauge Theories}},
  \href{https://doi.org/10.1016/0550-3213(83)90401-7}{\emph{Nucl. Phys. B}
  {\bfseries 228} (1983) 122}.

\bibitem{Ginsparg:1981bj}
P.H.~Ginsparg and K.G.~Wilson, \emph{{A Remnant of Chiral Symmetry on the
  Lattice}}, \href{https://doi.org/10.1103/PhysRevD.25.2649}{\emph{Phys. Rev.
  D} {\bfseries 25} (1982) 2649}.

\bibitem{Banks:1981ez}
T.~Banks and V.~Kaplunovsky, \emph{{Composite Massless Fermions in Strongly
  Coupled Lattice Gauge Theories}},
  \href{https://doi.org/10.1016/0550-3213(81)90203-0}{\emph{Nucl. Phys. B}
  {\bfseries 192} (1981) 270}.

\bibitem{Banks:1981nn}
T.~Banks and A.~Zaks, \emph{{On the Phase Structure of Vector-Like Gauge
  Theories with Massless Fermions}},
  \href{https://doi.org/10.1016/0550-3213(82)90035-9}{\emph{Nucl. Phys. B}
  {\bfseries 196} (1982) 189}.

\bibitem{Wilson:1984vkc}
K.G.~Wilson, \emph{{Monte Carlo Renormalization Group and the Three-Dimensional
  Ising Model}}, {\emph{NATO Sci. Ser. B} {\bfseries 115} (1984) 589}.

\bibitem{Golterman:1984ds}
M.F.L.~Golterman and J.~Smit, \emph{{Relation Between {QCD} Parameters on the
  Lattice and in the Continuum}},
  \href{https://doi.org/10.1016/0370-2693(84)90778-0}{\emph{Phys. Lett. B}
  {\bfseries 140} (1984) 392}.

\bibitem{Golterman:1984cy}
M.F.L.~Golterman and J.~Smit, \emph{{Selfenergy and Flavor Interpretation of
  Staggered Fermions}},
  \href{https://doi.org/10.1016/0550-3213(84)90424-3}{\emph{Nucl. Phys. B}
  {\bfseries 245} (1984) 61}.

\bibitem{Golterman:1984dn}
M.F.L.~Golterman and J.~Smit, \emph{{Lattice Baryons With Staggered Fermions}},
  \href{https://doi.org/10.1016/0550-3213(85)90138-5}{\emph{Nucl. Phys. B}
  {\bfseries 255} (1985) 328}.

\bibitem{Golterman:1985dz}
M.F.L.~Golterman, \emph{{Staggered Mesons}},
  \href{https://doi.org/10.1016/0550-3213(86)90383-4}{\emph{Nucl. Phys. B}
  {\bfseries 273} (1986) 663}.

\bibitem{Golterman:1986jf}
M.F.L.~Golterman, \emph{{Irreducible Representations of the Staggered Fermion
  Symmetry Group}},
  \href{https://doi.org/10.1016/0550-3213(86)90220-8}{\emph{Nucl. Phys. B}
  {\bfseries 278} (1986) 417}.

\bibitem{Daniel:1986zm}
D.~Daniel and T.D.~Kieu, \emph{{On the Flavor Interpretations of Staggered
  Fermions}}, \href{https://doi.org/10.1016/0370-2693(86)90334-5}{\emph{Phys.
  Lett. B} {\bfseries 175} (1986) 73}.

\bibitem{Kilcup:1986dg}
G.W.~Kilcup and S.R.~Sharpe, \emph{{A Tool Kit for Staggered Fermions}},
  \href{https://doi.org/10.1016/0550-3213(87)90285-9}{\emph{Nucl. Phys. B}
  {\bfseries 283} (1987) 493}.

\bibitem{Coste:1986qr}
A.~Coste, C.~Korthals~Altes and O.~Napoly, \emph{{Calculation of the Nonabelian
  Chiral Anomaly on the Lattice}},
  \href{https://doi.org/10.1016/0550-3213(87)90399-3}{\emph{Nucl. Phys. B}
  {\bfseries 289} (1987) 645}.

\bibitem{Konig:1985pw}
A.~Konig, K.H.~Mutter, K.~Schilling and J.~Smit, \emph{{Large Distance
  Propagators for Hadrons on a 56 X 16**3 Lattice}},
  \href{https://doi.org/10.1016/0370-2693(85)90393-4}{\emph{Phys. Lett. B}
  {\bfseries 157} (1985) 421}.

\bibitem{Bowler:1985ev}
K.C.~Bowler, F.~Gutbrod, P.~Hasenfratz, U.M.~Heller, F.~Karsch, R.D.~Kenway
  et~al., \emph{{The Beta Function and Potential at Beta = 6 and 6.3 in S(3)
  Gauge Theory}},
  \href{https://doi.org/10.1016/0370-2693(85)90298-9}{\emph{Phys. Lett. B}
  {\bfseries 163} (1985) 367}.

\bibitem{Luscher:1980ac}
M.~Luscher, \emph{{Symmetry Breaking Aspects of the Roughening Transition in
  Gauge Theories}},
  \href{https://doi.org/10.1016/0550-3213(81)90423-5}{\emph{Nucl. Phys. B}
  {\bfseries 180} (1981) 317}.

\bibitem{Kogut:1981yq}
J.B.~Kogut and D.K.~Sinclair, \emph{{On the Analyticity of the Off-axis String
  Tension and the Restoration of Rotational Symmetry in Lattice Systems}},
  \href{https://doi.org/10.1103/PhysRevD.24.1610}{\emph{Phys. Rev. D}
  {\bfseries 24} (1981) 1610}.

\bibitem{Luscher:1986pf}
M.~Luscher, \emph{{Volume Dependence of the Energy Spectrum in Massive Quantum
  Field Theories. 2. Scattering States}},
  \href{https://doi.org/doi:10.1007/BF01211097}{\emph{Commun. Math. Phys.}
  {\bfseries 105} (1986) 153}.

\bibitem{Luscher:1991cf}
M.~Luscher, \emph{{Signatures of unstable particles in finite volume}},
  \href{https://doi.org/10.1016/0550-3213(91)90584-K}{\emph{Nucl. Phys. B}
  {\bfseries 364} (1991) 237}.

\bibitem{Smit:1987wa}
J.~Smit and J.C.~Vink, \emph{{Neutral Pseudoscalar Masses in Lattice {QCD}}},
  \href{https://doi.org/10.1016/0550-3213(87)90034-4}{\emph{Nucl. Phys. B}
  {\bfseries 284} (1987) 234}.

\bibitem{Smit:1987zh}
J.~Smit and J.C.~Vink, \emph{{Renormalized Ward-takahashi Relations and
  Topological Susceptibility With Staggered Fermions}},
  \href{https://doi.org/10.1016/0550-3213(88)90354-9}{\emph{Nucl. Phys. B}
  {\bfseries 298} (1988) 557}.

\bibitem{Smit:1986fn}
J.~Smit and J.C.~Vink, \emph{{Remnants of the Index Theorem on the Lattice}},
  \href{https://doi.org/10.1016/0550-3213(87)90451-2}{\emph{Nucl. Phys. B}
  {\bfseries 286} (1987) 485}.

\bibitem{Hoek:1986nd}
J.~Hoek, M.~Teper and J.~Waterhouse, \emph{{Topological Fluctuations and
  Susceptibility in SU(3) Lattice Gauge Theory}},
  \href{https://doi.org/10.1016/0550-3213(87)90230-6}{\emph{Nucl. Phys. B}
  {\bfseries 288} (1987) 589}.

\bibitem{Berg:1981nw}
B.~Berg, \emph{{Dislocations and Topological Background in the Lattice O(3)
  $\sigma$ Model}},
  \href{https://doi.org/10.1016/0370-2693(81)90518-9}{\emph{Phys. Lett. B}
  {\bfseries 104} (1981) 475}.

\bibitem{Fucito:1980fh}
F.~Fucito, E.~Marinari, G.~Parisi and C.~Rebbi, \emph{{A Proposal for Monte
  Carlo Simulations of Fermionic Systems}},
  \href{https://doi.org/10.1016/0550-3213(81)90055-9}{\emph{Nucl. Phys. B}
  {\bfseries 180} (1981) 369}.

\bibitem{Swift:1984dg}
P.V.D.~Swift, \emph{{The Electroweak Theory on the Lattice}},
  \href{https://doi.org/10.1016/0370-2693(84)90350-2}{\emph{Phys. Lett. B}
  {\bfseries 145} (1984) 256}.

\bibitem{Smit:1985nu}
J.~Smit, \emph{{Fermions on a Lattice}}, {\emph{Acta Phys. Polon. B} {\bfseries
  17} (1986) 531}.

\bibitem{DHoker:1984mif}
E.~D'Hoker and E.~Farhi, \emph{{Decoupling a Fermion in the Standard
  Electroweak Theory}},
  \href{https://doi.org/10.1016/0550-3213(84)90587-X}{\emph{Nucl. Phys. B}
  {\bfseries 248} (1984) 77}.

\bibitem{Hands:1985ys}
S.J.~Hands and D.B.~Carpenter, \emph{{Lattice $\sigma$ Model and Fermion
  Doubling}}, \href{https://doi.org/10.1016/0550-3213(86)90093-3}{\emph{Nucl.
  Phys. B} {\bfseries 266} (1986) 285}.

\bibitem{Luscher:1987ay}
M.~Luscher and P.~Weisz, \emph{{Scaling Laws and Triviality Bounds in the
  Lattice phi**4 Theory. 1. One Component Model in the Symmetric Phase}},
  \href{https://doi.org/10.1016/0550-3213(87)90177-5}{\emph{Nucl. Phys. B}
  {\bfseries 290} (1987) 25}.

\bibitem{Luscher:1988gc}
M.~Luscher and P.~Weisz, \emph{{Is there a strong interaction sector in the
  Standard Lattice Higgs Model?}}, {\emph{Phys. Lett.} {\bfseries B212} (1988)
  472}.

\bibitem{Bock:1990tv}
W.~Bock, A.K.~De, K.~Jansen, J.~Jersak, T.~Neuhaus and J.~Smit, \emph{{Phase
  Diagram of a Lattice SU(2) X SU(2) Scalar Fermion Model With Naive and Wilson
  Fermions}}, \href{https://doi.org/10.1016/0550-3213(90)90689-B}{\emph{Nucl.
  Phys. B} {\bfseries 344} (1990) 207}.

\bibitem{Bock:1991kn}
W.~Bock, A.K.~De, C.~Frick, K.~Jansen and T.~Trappenberg, \emph{{Search for an
  upper bound of the renormalized Yukawa coupling in a lattice fermion Higgs
  model}}, \href{https://doi.org/10.1016/0550-3213(92)90692-5}{\emph{Nucl.
  Phys. B} {\bfseries 371} (1992) 683}.

\bibitem{Bock:1991bu}
W.~Bock, A.K.~De and J.~Smit, \emph{{Fermion masses at strong Wilson-Yukawa
  coupling in the symmetric phase}},
  \href{https://doi.org/10.1016/0550-3213(92)90551-L}{\emph{Nucl. Phys. B}
  {\bfseries 388} (1992) 243}.

\bibitem{Golterman:1991re}
M.F.L.~Golterman, D.N.~Petcher and J.~Smit, \emph{{Fermion interactions in
  models with strong Wilson-Yukawa couplings}},
  \href{https://doi.org/10.1016/0550-3213(92)90344-B}{\emph{Nucl. Phys. B}
  {\bfseries 370} (1992) 51}.

\bibitem{Aoki:1990yb}
S.~Aoki, I.-H.~Lee, J.~Shigemitsu and R.E.~Shrock, \emph{{Fermions in the
  Hypercharge Sector of the Standard Lattice Electroweak Theory: Global
  Limit}}, \href{https://doi.org/10.1016/0370-2693(90)91404-Y}{\emph{Phys.
  Lett. B} {\bfseries 243} (1990) 403}.

\bibitem{Aoki:1991es}
S.~Aoki, I.-H.~Lee and R.E.~Shrock, \emph{{Decoupling of doubler modes and
  tuning of fermion masses in the SU(2) x U(1)-Y lattice electroweak theory}},
  \href{https://doi.org/10.1103/PhysRevD.45.R13}{\emph{Phys. Rev. D} {\bfseries
  45} (1992) 13}.

\bibitem{Frick:1992ef}
C.~Frick, T.~Trappenberg, L.~Lin, G.~Munster, M.~Plagge, I.~Montvay et~al.,
  \emph{{Numerical simulation of heavy fermions in an SU(2)-L x SU(2)-R
  symmetric Yukawa model}},
  \href{https://doi.org/10.1016/0550-3213(93)90350-X}{\emph{Nucl. Phys. B}
  {\bfseries 397} (1993) 431}
  [\href{https://arxiv.org/abs/hep-lat/9207021}{{\ttfamily hep-lat/9207021}}].

\bibitem{Borrelli:1989kw}
A.~Borrelli, L.~Maiani, R.~Sisto, G.C.~Rossi and M.~Testa, \emph{{Neutrinos on
  the Lattice: The Regularization of a Chiral Gauge Theory}},
  \href{https://doi.org/10.1016/0550-3213(90)90041-B}{\emph{Nucl. Phys. B}
  {\bfseries 333} (1990) 335}.

\bibitem{Bock:1993qm}
W.~Bock, J.~Smit and J.C.~Vink, \emph{{Nongauge fixing approach to chiral gauge
  theories using staggered fermions}},
  \href{https://doi.org/10.1016/0550-3213(94)90326-3}{\emph{Nucl. Phys. B}
  {\bfseries 416} (1994) 645}
  [\href{https://arxiv.org/abs/hep-lat/9308009}{{\ttfamily hep-lat/9308009}}].

\bibitem{Banks:1992af}
T.~Banks and A.~Dabholkar, \emph{{Decoupling a fermion whose mass comes from a
  Yukawa coupling: Nonperturbative considerations}},
  \href{https://doi.org/10.1103/PhysRevD.46.4016}{\emph{Phys. Rev. D}
  {\bfseries 46} (1992) 4016}
  [\href{https://arxiv.org/abs/hep-lat/9204017}{{\ttfamily hep-lat/9204017}}].

\bibitem{Kaplan:1992bt}
D.B.~Kaplan, \emph{{A Method for simulating chiral fermions on the lattice}},
  \href{https://doi.org/10.1016/0370-2693(92)91112-M}{\emph{Phys. Lett. B}
  {\bfseries 288} (1992) 342}
  [\href{https://arxiv.org/abs/hep-lat/9206013}{{\ttfamily hep-lat/9206013}}].

\bibitem{Kaplan:1992sg}
D.B.~Kaplan, \emph{{Chiral fermions on the lattice}},
  \href{https://doi.org/10.1016/0920-5632(93)90282-B}{\emph{Nucl. Phys. B Proc.
  Suppl.} {\bfseries 30} (1993) 597}.

\bibitem{Luscher:1998pqa}
M.~Luscher, \emph{{Exact chiral symmetry on the lattice and the Ginsparg-Wilson
  relation}}, \href{https://doi.org/10.1016/S0370-2693(98)00423-7}{\emph{Phys.
  Lett. B} {\bfseries 428} (1998) 342}
  [\href{https://arxiv.org/abs/hep-lat/9802011}{{\ttfamily hep-lat/9802011}}].

\bibitem{Hamber:1983kx}
H.W.~Hamber, E.~Marinari, G.~Parisi and C.~Rebbi, \emph{{Numerical Simulations
  of Quantum Chromodynamics}},
  \href{https://doi.org/10.1016/0370-2693(83)91412-0}{\emph{Phys. Lett. B}
  {\bfseries 124} (1983) 99}.

\bibitem{Kronfeld:2007ek}
A.S.~Kronfeld, \emph{{Lattice Gauge Theory with Staggered Fermions: How, Where,
  and Why (Not)}}, \href{https://doi.org/10.22323/1.042.0016}{\emph{PoS}
  {\bfseries LATTICE 2007} (2007) 016}
  [\href{https://arxiv.org/abs/0711.0699}{{\ttfamily 0711.0699}}].

\bibitem{Wilson:1974evp}
K.G.~Wilson, \emph{{Quark Confinement}},  in \emph{{Marseille Colloquium on
  Lagrangian Field Theory}}, 1974,
  \href{https://www.osti.gov/biblio/4092791}{https://www.osti.gov/biblio/4092791}.

\bibitem{Korthals-Altes:1974xaz}
C.P.~Korthals~Altes, \emph{{Yang-Mills Fields on a Lattice and Dual
  Amplitudes}},  in \emph{{Marseille Colloquium on Lagrangian Field Theory}},
  1974,
  \href{https://www.osti.gov/biblio/4092791}{https://www.osti.gov/biblio/4092791}.

\end{thebibliography}\endgroup

\end{document}